\documentclass[superscriptaddress,amsmath,amssymb,aps,reprint,prb,twocolumn]{revtex4-2}

\usepackage{graphicx}% Include figure files
\usepackage{dcolumn}% Align table columns on decimal point
\usepackage{bm}% bold math
\usepackage{hyperref}% add hypertext capabilities
\usepackage[normalem]{ulem}
\hypersetup{
	citecolor = blue,
	colorlinks = true,
	urlcolor = blue
}
\usepackage{xcolor}
\usepackage{braket}

\newcommand{\sgn}{\mathrm{sgn}}

\begin{document}

\preprint{APS/123-QED}

\title{\textbf{Spin Peltier effect in graphene}}

\author{Xin Theng Lee}
\affiliation{Institute for Solid State Physics, University of Tokyo, Kashiwa 277-8581, Japan}

\author{Xin Hu}
\affiliation{Kavli Institute for Theoretical Sciences, University of Chinese Academy of Sciences, Beijing, 100190, China}

\author{Yuya Ominato}
\affiliation{Waseda Institute for Advanced Study, Waseda University, Shinjuku-ku, Tokyo 169-0051, Japan}

\author{Masahiro Tatsuno}
\affiliation{Institute for Solid State Physics, University of Tokyo, Kashiwa 277-8581, Japan}

\author{Takeo Kato}
\affiliation{Institute for Solid State Physics, University of Tokyo, Kashiwa 277-8581, Japan}

\author{Mamoru Matsuo}
\affiliation{Kavli Institute for Theoretical Sciences, University of Chinese Academy of Sciences, Beijing, 100190, China}
\affiliation{CAS Center for Excellence in Topological Quantum Computation, University of Chinese Academy of Sciences, Beijing 100190, China}
\affiliation{Advanced Science Research Center, Japan Atomic Energy Agency, Tokai, 319-1195, Japan}
\affiliation{RIKEN Center for Emergent Matter Science (CEMS), Wako, Saitama 351-0198, Japan}

\date{\today}

\begin{abstract}
In this work, we theoretically investigate the spin Peltier effect in a heterostructure composed of graphene and a ferromagnetic insulator (FI). Using a microscopic formalism based on characteristic scattering length at the FI, we analyze how spin accumulation in graphene give rise to temperature differences across the junction. We demonstrate that in the presence of an external magnetic field, the electronic spectrum of graphene is quantized into Landau Levels, which strongly modifies the available spin-flip scattering channels. In particular, the crossing of Landau levels significantly enhances the spin-flip scattering amplitude, leading to a pronounced amplification of the spin Peltier response. Our results suggest that measurements of the spin-induced temperature difference in graphene–ferromagnet heterostructures can serve as a sensitive probe of discrete electronic energy levels. More broadly, the present work provides a theoretical framework for understanding spin-driven thermal effects in hybrid systems combining Dirac materials and magnetic insulators.
\end{abstract}

\maketitle

\section{Introduction}
\label{sec:introduction}

The generation, manipulation, and detection of spin currents without relying on charge transport have become central themes in modern spintronics. 
Among the various mechanisms for spin current generation, thermal processes have attracted considerable interest over the past decade. 
The discovery of the spin Seebeck effect (SSE) revealed that a temperature gradient applied to a magnetic material can generate a spin current flowing along the direction of the thermal gradient~\cite{Uchida2008,Xiao2010,Adachi2011,Uchida2010,Jaworski2010,Masuda2024}. 
This phenomenon established the foundation of the field now known as spin caloritronics~\cite{Bauer2012,Uchida2021,Uchida2026}. 
Beyond its role as a spin-current generation mechanism, the SSE has also been utilized as a sensitive probe for thin films and low-dimensional systems~\cite{Ito2019,Kato2019,Li2019,Hirobe2019,Xing2022,Kikkawa2023,He2025,Kato2025}. 
This probe functionality originates from the fact that interfacial spin and energy transfer are governed by the elementary excitations and dynamical response functions of the material adjacent to the magnet.

A closely related probe technique is spin pumping (SP), in which coherent magnetization dynamics driven by ferromagnetic resonance injects spin angular momentum through an interface~\cite{Mizukami2001,Mizukami2001b,Mizukami2002,Tserkovnyak2002,Zutic2004,Tserkovnyak2005,Hellman2017}. 
While SP is now widely used as a method for spin injection into various materials, it also provides information about the dynamic spin susceptibility of the target material adjacent to the magnet. 
This property makes SP a sensitive probe for low-dimensional materials~\cite{Qiu2016,Han2020,Ominato2025}, for which traditional methods such as NMR~\cite{Harris2023} and neutron scattering~\cite{Wang2022} are often difficult to apply. 
In particular, because interfacial spin transfer is strongly affected by the electronic density of states, spin-current-based probes can detect discrete spectra and quantum oscillations in low-dimensional conductors.

These developments naturally raise the question of whether the spin Peltier effect, the reciprocal phenomenon of the SSE, can also be used as a spectroscopic probe under a magnetic field. 
In the spin Peltier effect, an injected spin current or spin accumulation induces a heat current or temperature difference through spin-energy conversion at an interface~\cite{Flipse2012,Flipse2014,Daimon2016,Ohnuma2017,Itoh2017,Sola2019}. 
Although the spin Peltier effect has been observed in magnetic heterostructures, its use as a probe of discrete electronic states in low-dimensional systems has not yet been explored. 
If the interfacial spin-flip scattering processes responsible for the spin Peltier effect are controlled by the electronic spectrum, the induced temperature response should carry direct information about the available spin-flip channels.

\begin{figure}[tb]
    \centering
    \includegraphics[width=0.9\linewidth]{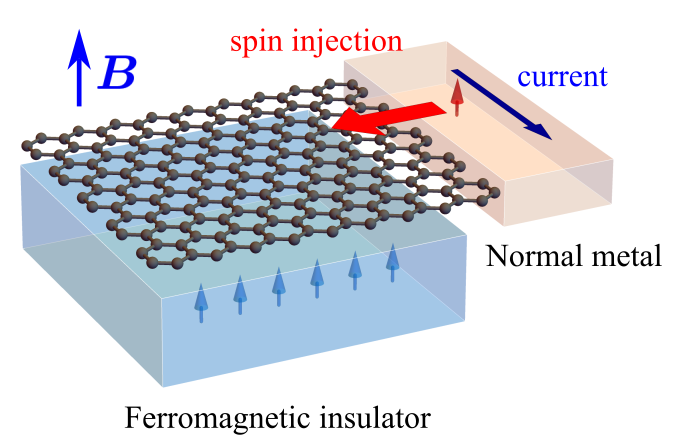}
    \caption{Schematic illustration of a graphene/ferromagnetic-insulator (FI) heterostructure. An external magnetic field $\bm B$ is applied perpendicular to the graphene plane.
    The spin Peltier effect is induced by spin accumulation in graphene, which can be, for example, driven by the spin Hall effect in an adjacent normal metal.
    }
    \label{fig:model}
\end{figure}

In this paper, we study the spin Peltier effect in a bilayer system composed of a ferromagnetic insulator (FI) and graphene under a perpendicular magnetic field (see Fig.~\ref{fig:model}). 
Graphene provides an ideal platform for this purpose because its Landau levels are well resolved at low temperatures and can be tuned by the magnetic field and carrier density. 
For graphene/FI systems, quantum oscillations arising from Landau quantization have recently been discussed in the context of the SSE and SP~\cite{Ominato2020,Hu2024}. 
Here, we address the reciprocal spin Peltier response induced by spin accumulation in graphene, which can be generated, for example, by the spin Hall effect in an adjacent metal as shown in Fig.~\ref{fig:model}. 
Based on a microscopic model of interfacial exchange coupling, we calculate the temperature difference generated by the spin accumulation in graphene. 
We show that Landau-level crossings strongly enhance the spin-flip scattering channels and produce pronounced oscillations of the induced temperature difference as a function of inverse magnetic field. 
We also include the effect of phonon heat leakage, which reduces the absolute magnitude of the temperature response but does not obscure the oscillatory structure. 
Our results demonstrate that the spin Peltier effect in graphene/FI heterostructures can provide a novel thermal probe of discrete electronic states in solid-state materials.

The remainder of the paper is organized as follows. 
Section~\ref{sec:model} introduces a microscopic model for graphene/FI junctions. Section~\ref{sec:formulation} derives the spin and energy currents generated by interfacial spin-flip scattering. 
From the balance conditions on these currents, we also formulate the spin-induced temperature difference across the junction in the presence of phonon heat leakage.
Section~\ref{sec:results} presents the magnetic-field dependence of the spin Peltier response, as well as an estimate of the temperature difference. 
Finally, we summarize our results in Section~\ref{sec:summary}.

\section{Model} \label{sec:model}

\begin{figure}
\centering
\includegraphics[width=0.9\linewidth]{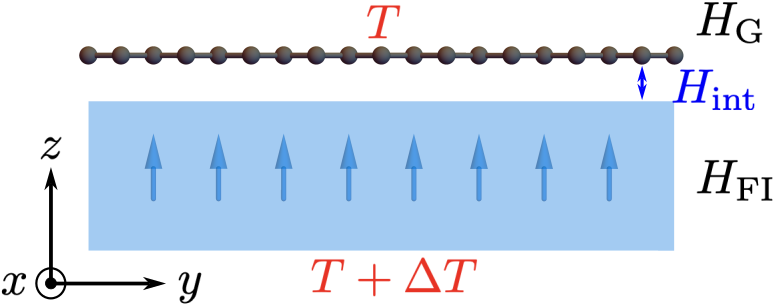}
\caption{Schematic illustration of the FI-grphene junction.}
\label{fig:zerospincurrent}
\end{figure}

In this section, we consider a hybrid system consisting of monolayer graphene coupled to a ferromagnetic insulator (FI) as shown in Fig.~\ref{fig:zerospincurrent}.
The total Hamiltonian is given by
\begin{align}
    H = H_{\rm G} + H_{\rm FI} + H_{\rm int}.
\end{align}
Here, $H_{\rm G}$, $H_{\rm FI}$, and $H_{\rm int}$ describe graphene, the magnetic insulator, and the interfacial exchange interaction, respectively.
In the following, we present explicit forms of these Hamiltonians.

\subsection{Graphene}

Graphene is described by a tight-binding Hamiltonian on a honeycomb lattice,
\begin{align}
    H_{\rm G} = -t \sum_{\langle i,j \rangle} \left( a_i^\dagger b_j + \mathrm{h.c.} \right),
\end{align}
where $t$ is the hopping amplitude, $\langle i,j \rangle$ denotes nearest-neighbor pairs, and $a_i$ ($b_j$) is the annhilation operator for electrons on sublattice A (B).
Introducing the Fourier transforms, $a_i = N^{-1/2} \sum_{\bm k} a_{\bm k} e^{i \bm k \cdot \bm R_i}$ and $b_j = N^{-1/2} \sum_{\bm k} b_{\bm k} e^{i \bm k \cdot \bm R_j}$, where $N$ is the number of unit cells, the Hamiltonian can be written as
\begin{align}
    H_{\rm G} &= \sum_{\bm k} 
    \begin{pmatrix}
        a_{\bm k}^\dagger & b_{\bm k}^\dagger
    \end{pmatrix}
    H(\bm k)
    \begin{pmatrix}
        a_{\bm k} \\ b_{\bm k}
    \end{pmatrix}, \\
    H(\bm k) &= 
    \begin{pmatrix}
        0 & -t\gamma_{\bm k} \\
        -t\gamma_{\bm k}^\ast & 0
    \end{pmatrix},
\end{align}
where $\gamma_{\bm k} = \sum_{n=1}^{3} e^{i \bm k \cdot \bm \delta_n}$.

The low-energy Hamiltonian near the Dirac points $\bm K^\xi$ with valley index $\xi = \pm 1$ is given by
\begin{align}
    H^\xi(\bm q) = \hbar v_F
    \begin{pmatrix}
        0 & \xi q_x - i q_y \\
        \xi q_x + i q_y & 0
    \end{pmatrix},
\end{align}
where ${\bm q} = \bm k - \bm K^\xi$ is a small deviation of the wave vector from the Dirac points, and $v_F = 3ta/(2\hbar)$ is the Fermi velocity.

A perpendicular magnetic field $B$ is introduced via a minimum coupling $\hbar \bm q \rightarrow \bm \pi = \hbar \bm q + e \bm A$, where $e>0$ is the electron charge.
Employing the Landau gauge $\bm A = (0,Bx,0)$, the eigenstates for the $K^+$ valley become 
\begin{align}
    \Phi_{n q_y +}({\bm r}) = \left\{ \begin{array}{ll}
    \begin{pmatrix}
        0 \\ \phi_{0,q_y}
    \end{pmatrix}, & (n=0), \\
    \frac{1}{\sqrt{2}}  \begin{pmatrix}
        \sgn(n) \phi_{|n|-1,q_y} \\ \phi_{|n|,q_y}
    \end{pmatrix}, & (n\ne 0) ,
    \end{array} \right. \label{Phiplus}
\end{align}
and for the $K^-$ valley  
\begin{align}
    \Phi_{n q_y -}({\bm r}) = \left\{ \begin{array}{ll}
    \begin{pmatrix}
        \phi_{0,q_y} \\ 0
    \end{pmatrix}, & (n=0), \\
    \frac{1}{\sqrt{2}}
    \begin{pmatrix}
         \phi_{|n|,q_y} \\ -\sgn(n) \phi_{|n|-1,q_y}
    \end{pmatrix}, & (n\ne 0),
    \end{array} \right.  \label{Phiminus}
\end{align}
where $n \in \mathbb{Z}$ is the Landau-level index, $q_y$ is the $y$ component of the wave vector, and $\phi_{n,q_y}$ is a wave function defined by
\begin{align}
    \phi_{n,q_y}(\bm r) &= \frac{1}{\sqrt{L_y}} e^{iq_y y} f_n(x - x_0(q_y)) , \label{phinky}\\
    f_n(x) &= N_ne^{-(x/\ell_B)^2/2}H_n(x/\ell_B), \label{phinky2} \\
    x_0(q_y) &= - \frac{\hbar q_y}{eB}, \label{phinky3}
\end{align}
with the Hermite polynomial $H_n(x/\ell_B)$ and the normalization constant $N_n = (\pi^{1/2}2^n n!\ell_B)^{-1/2}$.
The corresponding eigenenergy is independent of $q_y$ and is given by
\begin{align} 
        \varepsilon_{n} &= \mathrm{sgn}(n)\, \hbar \omega_c \sqrt{|n|},
\end{align}
where $\omega_c = \sqrt{2} v_F/\ell_B$ and $\ell_B = \sqrt{\hbar/(eB)}$ is the magnetic length. For a detailed derivation, see Appendix~\ref{app:LandauLevels}.

By considering the spin degree of freedom, the electron energy is modified as
\begin{align}
     \varepsilon_{n\sigma} = \sgn(n) \hbar \omega_c\sqrt{|n|} - g \mu_{\rm B} \sigma B/2,
\end{align}
where $g$ is the $g$-factor, $\mu_{\rm B}$ is the Bohr magneton, and $\sigma=\pm 1$ denotes the $z$-component of the electron spin.
Using these eigenenergies, the Hamiltonian of graphene is given by
\begin{align}
{\cal H}_{\rm G} & = H_{\rm G} - \mu N \notag \\
&= \sum_{n,q_y,\xi,\sigma} (\varepsilon_{n\sigma} - \mu)c_{n q_y\xi  \sigma}^\dagger c_{n q_y \xi \sigma} ,
\end{align}
where $N=\sum_{n,q_y,\xi,\sigma} c_{n q_y \xi \sigma}^\dagger c_{n q_y \xi \sigma}$ is the electron number operator, $\mu$ is the chemical potential of graphene, and $c_{n q_y \xi \sigma}$ ($c_{n q_y \xi \sigma}^\dagger$) is an annihilation (creation) operator of electrons under the external magnetic field.

\subsection{Ferromagnetic Insulator}
\label{sec:ferroinsulator}

We consider a three-dimensional ferromagnetic insulator on a lattice described by the Heisenberg Hamiltonian
\begin{align}
    H_{\rm FI} = - J \sum_{\langle {\bm R},{\bm R}'\rangle} \bm S_{\bm R} \cdot {\bm S}_{{\bm R}'}  - \hbar \gamma B \sum_{\bm R} \, S_{\bm R}^z ,
\end{align}
where $\bm S_{\bm R}$ denotes the localized spin operator at position $\bm R$, $\langle {\bm R},{\bm R}'\rangle$, $J>0$ is the exchange coupling constant, $\gamma$ is the gyromagnetic ratio, and $B$ is the external magnetic field applied along the $z$ direction.

We employ the standard spin-wave approximation based on the Holstein-Primakoff transformation~\cite{HolsteinPrimakoff1940}:
$S_{\bm{R}}^z = S_0 - b_{\bm{R}}^{\dagger}b_{\bm{R}}$,
$S_{\bm{R}}^+ \approx (2S_0)^{1/2} b_{\bm{R}}$, and
$S_{\bm{R}}^- = (S_{\bm{R}}^+)^\dagger$,
where $S_0$ is the spin magnitude and $b_{\bm R}$ is the magnon annihilation operator.
Using the Fourier transform of the magnon operator
\begin{align}
b_{\bm R}= \frac{1}{N_{\rm FI}^{1/2}} \sum_{\bm k} e^{-i{\bm k}\cdot {\bm R}} b_{\bm k},
\end{align}
with the number of unit cells $N_{\rm FI}$, the Hamiltonian is rewritten as
\begin{align}
H_{\rm FI} \simeq \sum_{\bm k} \hbar \omega_{\bm k} b_{\bm k}^\dagger b_{\bm k} ,
\end{align}
where $\hbar \omega_{\bm k}={\cal D}|\bm k|^2+\hbar \gamma B$ is the magnon dispersion relation and ${\cal D}=JS_0a_{\rm FI}^2$ is the spin stiffness.

\subsection{Interfacial exchange coupling}
\label{sec:InterfaceModel}

The exchange coupling at the graphene/FI interface is given by
\begin{align}
H_{\rm int} = -\sum_{{\bm r}\alpha} J_{\bm r \alpha} \bm S_{\bm r} \cdot \bm s_{{\bm r} \alpha}, 
\end{align}
where $\bm S_{\bm r} = \bm S_{{\bm R}={\bm r}}$ is the ferromagnetic local spin operator, and $\bm s_{{\bm r}\alpha}$ is the spin operator for a sublattice $\alpha$ defined by
\begin{align}
\bm s_{{\bm r}\alpha} &= a_{\rm G}^2 \sum_{n,q_y,\xi,\sigma,\sigma'} (\Phi_{n q_y \xi}({\bm r})^*)_\alpha c_{n q_y \xi \sigma}^\dagger ({\bm \sigma}/2)_{\sigma \sigma'} \notag \\
& \hspace{15mm} \times (\Phi_{n q_y \xi}({\bm r}))_\alpha c_{n q_y \xi \sigma'} .
\end{align}
Here, $\alpha$ denotes the sublattice of graphene and ${\bm \sigma} = (\sigma_x, \sigma_y. \sigma_z)$ are the Pauli matrices.
Using the spin-wave approximation, the Hamiltonian is modified as
\begin{align}
H_{\rm int} &\simeq H_{\rm T} + H_{\rm Z} , \\
H_{\rm T} &= -\frac{1}{2} \sum_{{\bm r}\alpha}  J_{{\bm r} \alpha} \sqrt{2S_0} (b_{\bm r}s^-_{{\bm r}\alpha} + b^\dagger_{\bm r}s^+_{{\bm r}\alpha}), \\
H_{\rm Z} &= - \sum_{{\bm r}\alpha} J_{{\bm r}\alpha} S_0 s_{{\bm r}\alpha}^z ,
\end{align}
where $s^\pm_{{\bm r}\alpha} = s^x_{{\bm r}\alpha} \pm i s^y_{{\bm r}\alpha}$ and the two dominant terms are left in the expansion of $1/S_0$.
The Hamiltonian $H_{\rm Z}$ describes the exchange bias due to the interfacial coupling and can be incorporated by replacing the electron eigenstates with 
\begin{align}
     \varepsilon_{n \sigma} = \sgn(n) \hbar \omega_c\sqrt{|n|} - J_0 S_0 \sigma/2,
\end{align}
where $J_0$ is the average of the exchange interaction $J_{{\bm r}\alpha}$ per site.
For simplicity, we have assumed that the exchange bias is sufficiently larger than the Zeeman term due to the electron spins.

\section{Formulation}
\label{sec:formulation}

In this section, we formulate the temperature difference induced by the spin Peltier effect.
Regarding ${\cal H}_0 = {\cal H}_{\rm G} + H_{\rm FI} + H_{\rm Z}$ as the unperturbed Hamiltonian, we first calculate the spin current using second-order perturbation with respect to $H_{\rm T}$ within the Keldysh formalism.
Next, we derive the temperature difference from the steady-state condition of the junction system.

\subsection{Local spin susceptibility}
\label{sec:spinsusceptibility}

Before showing the perturbation calculation, we first calculate the imaginary part of the local spin susceptibilities for isolated graphene and FI, which will be used for calculating the spin current. 
This is defined by 
\begin{align}
\chi_{\rm loc}^R(\omega) &= \int dt \, e^{i\omega t} \chi^R_{\rm loc}(t), \\
\chi_{\rm loc}^R(t) &=
-\frac{i}{\hbar} \theta(t) \sum_{\alpha} \langle [s_{{\bm r}\alpha}^+(t),s_{{\bm r}\alpha}^-(0)] \rangle_0 , 
\end{align}
where $\langle \cdots \rangle_0 = {\rm Tr} \, [e^{-\beta {\cal H}_{\rm G}} \cdots]/{\rm Tr} \, [e^{-\beta {\cal H}_{\rm G}}]$ denotes the thermal average for graphene, $a_{\rm G}^2$ denotes the microscopic area associated with one carbon site, $\theta(x)$ is the Heaviside step function, and $s_{\bm r}^\pm(t) = e^{i{\cal H}_{\rm G}t/\hbar} s_{\bm r}^\pm e^{-i{\cal H}_{\rm G}t/\hbar}$.
We note that both $\chi_{\rm loc}(\omega) $ and $\chi_{\rm loc}(t)$ are independent of the position ${\bm r}$ due to the translational invariance of the continuum Hamiltonian of graphene.
In the following, we abbreviate the index of eigenstates in graphene $(n,q_y,\xi,\sigma)$ as $j$.
The local spin susceptibility is expressed with the eigenfunction of the graphene Eqs.~\eqref{phinky}-\eqref{phinky3} as
\begin{align}\label{eq:sumformlocalsup}
    \text{Im}\, \chi_{\rm loc}^R(\omega)  &=  \pi a_{\rm G}^4 \sum_{j,j',\alpha}|(\Phi_j(\bm r)^*)_\alpha (\Phi_{j'}(\bm r))_\alpha|^2 |s_{\sigma \sigma'}^+|^2 \notag \\
    & \hspace{-5mm} \times \delta (\hbar \omega + \varepsilon_j - \varepsilon_{j'})[f_{\rm F}(\varepsilon_j) - f_{\rm F}(\varepsilon_j + \hbar \omega)].
\end{align}
where $f_{\rm F}(\varepsilon) = (e^{\beta \varepsilon}+1)^{-1}$ is the Fermi-Dirac distribution.
A straightforward calculation gives~\cite{Hu2024}
\begin{align}\label{eq:localspinsuscep}
    &{\rm Im} \, \chi_{\rm loc}^R(\omega) \notag\\
    & = \frac{\pi a_{\rm G}^4}{2} \int d\varepsilon [f_{\rm F}(\varepsilon) - f_{\rm F}(\varepsilon + \hbar \omega)] D_+^{\rm LL}(\varepsilon)D_-^{\rm LL}(\varepsilon + \hbar \omega),
\end{align}
where 
\begin{align}\label{llDOS}
    D_{\sigma}^{\rm LL}(\varepsilon) = \frac{1}{2\pi \ell_B^2} \sum_{n} \delta(\varepsilon - \varepsilon_{n \sigma}),
\end{align}
is the density of states per unit area in graphene. 
In realistic systems, Landau levels are broadened by impurity scattering or randomness. 
This can be captured by replacing $\delta(\varepsilon - \varepsilon_{n \sigma})$ with a Lorentzian function as
\begin{align}
    \delta(\varepsilon - \varepsilon_{n \sigma}) \to \frac{1}{\pi}\frac{\Gamma}{(\varepsilon - \varepsilon_{n \sigma})^2 + \Gamma^2},
\end{align}
where $\Gamma$ denotes the scattering rate.

The dynamic spin susceptibility of the FI is defined as 
\begin{align}
G_{\rm loc}^R(\omega) &= \int dt \, e^{i\omega t} G_{\rm loc}(t), \\
G_{\rm loc}^R(t) &=
- \frac{i}{\hbar} \theta(t) \langle [S_{{\bm r}}^+(t),S_{{\bm r}}^-(0)] \rangle_0.
\end{align}
We note that both $G_{\rm loc}^R(\omega) $ and $G_{\rm loc}^R(t)$ are independent of the position ${\bm R}$ due to the translational invariance of the FI.
Using the spin-wave approximation, the imaginary part of the dynamic spin susceptibility is calculated as
\begin{align}
{\rm Im}\, G_{\rm loc}^R(\omega) = - \frac{{J}^{-3/2}} {2\pi \sqrt{S_0}} \sqrt{\hbar \omega - \hbar \gamma B} \,
\theta(\hbar\omega-\hbar\gamma B).
\end{align}
We note that $-{\rm Im}\, G_{\rm loc}^R(\omega)/2\pi S_0$ corresponds to the magnon density of states per FI unit cell. 

\subsection{Spin and energy currents}

The tunneling spin current operator is defined by
\begin{align}
\hat{I}_{\rm S} &= \frac{i}{\hbar} [s_z^{\rm tot} , H] \notag \\
&=\frac{i}{2} \sum_{{\bm r}\alpha} (J_{{\bm r}\alpha} S_{\bm{r}}^{+} s_{{\bm r}\alpha}^- - {\rm h.c.}),
\end{align}
where $s_z^{\rm tot} = \hbar \sum_{{\bm r}\alpha} s^z_{{\bm r}\alpha}$ is the total spin in graphene.
Within the Keldysh formalism, we expand the spin current up to second-order in $H_{\rm T}$ and take a random average for the interfacial exchange coupling using
\begin{align}
    \langle J_{\bm r \alpha} J^\ast_{\bm r'\alpha'} \rangle_{\rm ave} &= J_{\rm int}^2 \delta_{{\bm r},{\bm r}'} \delta_{\alpha,\alpha'}, \label{eq:Jav}
\end{align} 
where $J_{\rm int}$ represents the strength of the interfacial exchange interaction, $a^2$ is the unit-cell area of graphene, and $N_{\rm int}$ is the number of interfacial exchange bonds~\cite{Ominato2025}.
Then, the spin current per unit area is given as
\begin{align}
    & j_{\rm S} = \langle \hat{I}_{\rm S}
    \rangle/{\cal A} \notag \\
    & = A  \int \frac{d (\hbar \omega)}{2 \pi} \, \text{Im}\, \chi_{\text{loc}}^R(\omega) \, \text{Im}\, (-G_{\rm loc}^R(\omega))[f_{\rm FI}(\omega)-f_{\rm G}(\omega)] , \label{eq:spincurrent}
\end{align}
where $A = J_{\textrm{int}}^2 n_{\rm int}$, $n_{\rm int}$ is the number of the interfacial bonds per unit area, and $f_{\rm G}(\omega)$ and $f_{\rm FI}(\omega)$ are nonequilibrium distribution functions for graphene and FI, respectively.
For detailed derivation, see Appendices~\ref{appex:nonequiv} and \ref{appex:derivationofSpinCurrent}.

In a similar way, the energy current per unit area is calculated as~\cite{Ohnuma2017}
\begin{align}
    j_{\rm E} = &A \int \frac{d (\hbar \omega)}{2 \pi} \hbar \omega \, \text{Im}\, \chi_{\text{loc}}^R(\omega) \, \text{Im}\, (-G_{\rm loc}^R(\omega)) \notag \\
    & \times [f_{\rm FI}(\omega)-f_{\rm G}(\omega) ] . \label{eq:hearcurrent}
\end{align}
Compared with the spin current in Eq.~\eqref{eq:spincurrent}, the factor $\hbar \omega$ is added to the integrand~\footnote{The heat current, which is defined by $I_{\rm Q} = I_{\rm E} - \Delta \mu \langle I_{\rm S}\rangle$ with the spin chemical potential difference $\Delta \mu = \mu_{\rm G} - \mu_{\rm FI}$, is approximately equal to the energy current because the condition $I_{\rm E} \gg \mu_G \langle I_{\rm S}\rangle$ usually holds. Indeed, this approximation is implicitly employed in almost all of the theoretical studies of spin caloritronics.}.
In the present junction system, the spin Peltier effect is induced by spin accumulation in graphene driven by external sources.

\subsection{Linear response}

The spin accumulation in graphene, which drives the temperature difference across the junction, is incorporated through the change of the density matrix $e^{-\beta {\cal H}_{\rm G}} \rightarrow e^{-\beta ({\cal H}_{\rm G}+\Delta {\cal H}_{\rm G})}$, where 
\begin{align}
\Delta {\cal H}_{\rm G} = - \frac{\mu_{\rm G}}{2} \sum_{n,q_y,\xi,\sigma=\pm 1} \sigma  c^\dagger_{n q_y \xi \sigma} c_{n q_y \xi \sigma}, 
\end{align}
where $\mu_{\rm G}= \mu_\uparrow - \mu_\downarrow$ is the spin accumulation in graphene.
This modifies the nonequilibrium distribution function $f^{\rm G}(\omega)$.
By a similar calculation as in deriving the fluctuation-dissipation theorem~\cite{Kato2020}, we obtain
\begin{align}
f_{\rm G}(\omega) = f_{\rm B}(\hbar \omega + \mu_{\rm G}) ,
\end{align}
where $f_{\rm B}(\epsilon) = (e^{\beta \epsilon}-1)^{-1}$ is the Bose distribution function.

In a similar way, we can incorporate magnon accumulation in the FI by the change of the density matrix $e^{-\beta H_{\rm FI}}\rightarrow e^{-\beta ({\cal H}_{\rm FI}+\mu_{\rm FI} N_{\rm FI})}$, where $N_{\rm FI} = \sum_{\bm k} b_{\bm k}^\dagger b_{\bm k}$ is the magnon number operator.
The plus sign appears because a magnon carries spin angular momentum opposite
to the magnetization direction; thus $\mu_{\rm FI}$ is the spin chemical
potential, not the magnon-number chemical potential.
This gives the nonequilibrium distribution function
\begin{align}
f_{\rm FI}(\omega) = f_{\rm B}(\hbar \omega + \mu_{\rm FI}) .
\end{align}

In this work, we discuss the spin Peltier effect within the linear response.
We can expand the difference between the two distribution functions, $f_{\rm G}(\omega)$ and $f_{\rm FI}(\omega)$, to the first order of $ \mu_{\rm G}$ and temperature difference $\Delta T = T_{\rm FI}-T_{\rm G}$ as 
\begin{align}\label{eq:fermidiff}
    & f_{\rm FI}(\omega) - f_{\rm G}(\omega) \notag \\
    &\approx \frac{\beta }{4\sinh^2(\beta \hbar \omega/2)} (\mu_{\rm G}-\mu_{\rm FI}) +\frac{k_B\beta^2 \hbar \omega}{4\sinh^2(\beta \hbar \omega/2)}\Delta T ,
\end{align}
where $\beta=1/k_{\rm B}T$ is the inverse temperature and $T = T_{\rm G}$. The spin and energy current satisfies the Onsager's reciprocal relations
\begin{align}
\left( \begin{array}{c} 
j_{\rm S} \\ j_{\rm E} \end{array} \right) 
= \left( \begin{array}{cc} 
L_{\rm SS} & L_{\rm SE} \\
L_{\rm ES} & L_{\rm EE} 
\end{array}\right)
\left( \begin{array}{c} 
\mu_{\rm G} - \mu_{\rm FI} \\ \Delta T/T \end{array} \right) .
\label{eq:linearrelations}
\end{align}
with the matrix coefficients given as
\begin{align}
    L_{\rm SS} &= A\int\frac{d(\hbar\omega)}{2\pi}{\cal F}(\omega),\\
    L_{\rm SE} &= L_{\rm ES}
    = A\int\frac{d(\hbar\omega)}{2\pi}\hbar\omega\,{\cal F}(\omega),\\
    L_{\rm EE} &= A\int\frac{d(\hbar\omega)}{2\pi}(\hbar\omega)^2{\cal F}(\omega),
\end{align}
where
\begin{align}
{\cal F}(\omega)=\,
{\rm Im}\,\chi_{\rm loc}^R(\omega,B)
\, {\rm Im}\,(-G_{\rm loc}^R(\omega))
\frac{\beta}{4\sinh^2(\beta\hbar\omega/2)}.
\end{align}

\subsection{Induced temperature difference}
\label{tempdiff}

For further calculations, we need to determine the temperature difference $\Delta T$ and the chemical potential $\mu_{\rm FI}$ from a fixed $\mu_{\rm G}$.
If the spin relaxation in the FI is weak, the FI spin chemical potential $\mu_{\rm FI}$ tends to saturate up to $\mu_{\rm G}$, leading to the vanishing of the spin current.
Therefore, to obtain a large spin current across the junction, we assume strong spin relaxation in the FI in this work, for which $\mu_{\rm FI} \simeq 0$ (for a detailed discussion, see Appendix~\ref{app:mus}). 
For this situation, the electron spin is not a conserved quantity, and the steady-state condition cannot be given by the spin current.
On the other hand, the steady-state condition can be expressed as a balance condition between the heat current carried by spins and the thermal backflowing heat current by phonons:
\begin{align}
j_E + G_{\rm ph} \Delta T = 0.
\end{align}
Here, $G_{\rm ph}$ is the phonon thermal conductance of the graphene/FI interface.
Using Eq.~\eqref{eq:linearrelations}, the observable temperature response becomes
\begin{align}\label{eq:generaltempdiff}
    \frac{\Delta T}{\mu_{\rm G}}
    =
    -\frac{T L_{ES}}{L_{EE}+G_{\rm ph}T}.
\end{align}
This is the main spin Peltier coefficient discussed in this work.

For usual experimental conditions, the linear response coefficient of the heat current with respect to the temperature, $L_{EE}$, is sufficiently small compared with $G_{\rm ph}T$, since the former is determined by small interfacial exchange coupling, while the latter is determined by the phonon transmission across the junction.
Then, the temperature difference is approximately given by
\begin{align}\label{eq:generaltempdiff2}
    \frac{\Delta T}{\mu_{\rm G}}
    \simeq
    -\frac{L_{ES}}{G_{\rm ph}}.
\end{align}
Here, the phonon thermal conductance reduces the absolute magnitude of the temperature response. 
However, it is expected to be a smooth function of magnetic field and therefore does not contribute to the quantum oscillations associated with Landau quantization.
Thus, the oscillatory component of $\Delta T(B)$ is a fingerprint of the interfacial spin Peltier response mediated by interfacial spin--magnon conversion.

\section{Results}
\label{sec:results}

\begin{figure}
    \centering
    \includegraphics[width=1.0\linewidth]{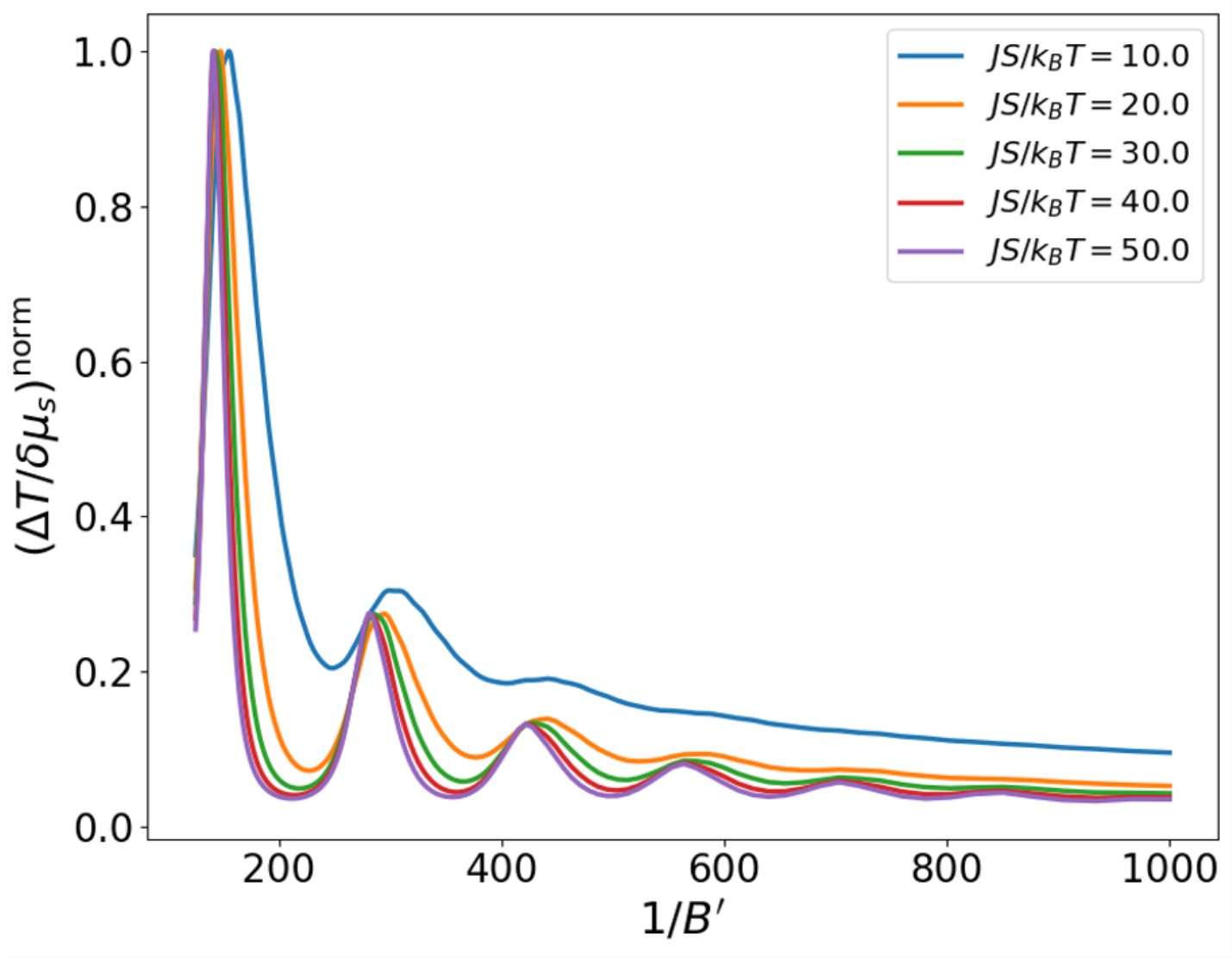}
    \caption{Quantum oscillation of the normalized temperature difference $(\Delta T/\delta\mu_s)^{\rm norm}=(\Delta T/\delta\mu_s)/\max(\Delta T/\delta\mu_s)$ plotted as a function of the inverse dimensionless magnetic field $1/B'$. The lines correspond to different values of the temperature-control parameter $JS/k_BT=10,20,30,40,50$. The Landau-level broadening is fixed at $\Gamma/JS=1$, with chemical potential $\mu/JS=1$. The oscillatory peaks originate from successive crossings of graphene Landau levels with the thermally active electronic window.}
    \label{fig3}
\end{figure}

In this section, we discuss the results obtained from our analysis. 
To analyze the temperature dependence of the magnon-mediated quantum oscillation, we rewrite the response $\Delta T/ \delta \mu_s$ in a dimensionless form by measuring the electronic energy and the magnon energy in units of the exchange scale $JS$. We introduce $\Gamma' = \Gamma/JS$, $B' = \hbar \gamma B /JS$, and $JS/k_BT$ is the control parameter. In this representation, the Landau-level energies become $\varepsilon_{n\sigma}' =  \varepsilon_{n\sigma}/JS$, while the thermal magnon factor is controlled by $1/\sinh^2(JS \Omega/2k_bT)$. Therefore, $JS/k_BT$ directly characterizes the competition between Landau-level resolution and thermal activation of magnon-mediated transitions. Large $JS/k_BT$ corresponds to low temperature, where the electronic Landau levels are sharp but magnon-assisted processes are thermally suppressed, whereas small $JS/k_BT$ enhances magnon occupation but broadens the electronic thermal window. 
The line plot shown in Fig.~\ref{fig3} is calculated from the
dimensionless expression for the normalized temperature difference
$(\Delta T/\delta\mu_s)^{\rm norm}$ as a function of the inverse
dimensionless magnetic field $1/B'$. Here $JS/k_BT$ is used as the main control
parameter for the visibility of the quantum oscillations, while $\mu/JS$
and $\Gamma/JS$ are fixed to unity. Since $JS/k_BT$ characterizes the ratio
between the magnetic energy scale and the thermal energy, increasing the
temperature at fixed $JS$ effectively enhances thermal smearing. As a
result, the discrete Landau-level structure becomes less resolved, and the
oscillatory component of $(\Delta T/\delta\mu_s)^{\rm norm}$ is
progressively suppressed. This behavior indicates that the oscillations
originate from Landau quantization and that their visibility is limited by
thermal broadening.

\begin{figure}
    \centering
\includegraphics[width=1.0\linewidth]{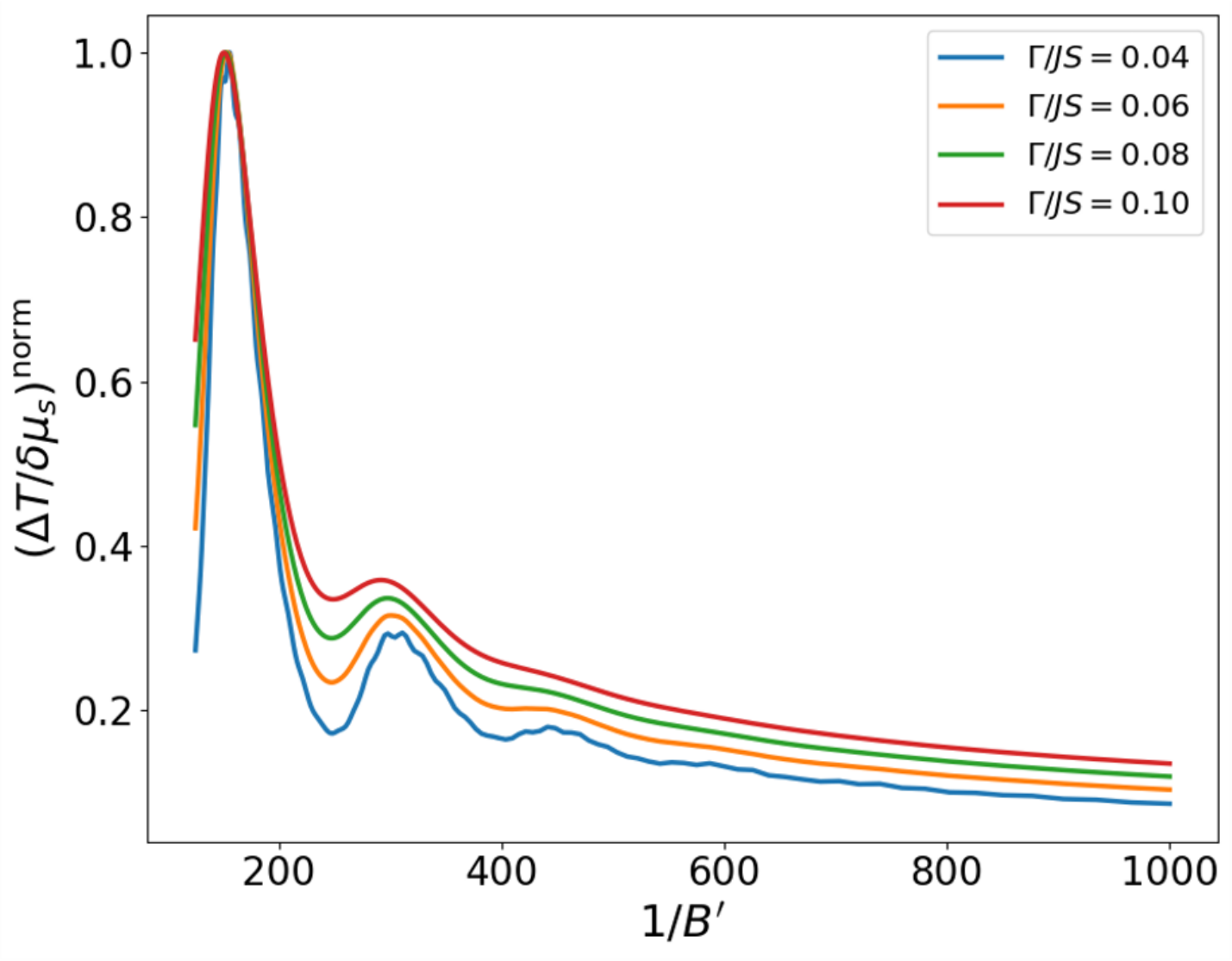}
    \caption{Quantum oscillation of the normalized temperature difference $(\Delta T/\delta\mu_s)^{\rm norm}=(\Delta T/\delta\mu_s)/\max(\Delta T/\delta\mu_s)$ plotted as a function of the inverse dimensionless magnetic field $1/B'$. The lines correspond to different values of the Landau-level broadening parameter $\Gamma/JS=0.04,0.06,0.08,0.1$. The temperature control parameter is fixed at $JS/k_BT=10$, with chemical potential $\mu/JS=1$. The oscillatory peaks originate from successive crossings of graphene Landau levels with the thermally active electronic window.}
    \label{fig4}
\end{figure}

Fig.~\ref{fig4} demonstrates the role of Landau-level broadening in the
normalized spin-Peltier coefficient $(\Delta T/\delta\mu_s)^{\rm norm}$.
Here $JS/k_B T=10$ is fixed, while varying $\Gamma/JS$. The temperature difference shows
a pronounced peak followed by a series of weaker oscillations as a function
of $1/B'$, reflecting the successive passage of Landau levels through the
thermally active energy window. For small $\Gamma$, the Landau levels remain
well resolved, giving rise to clear quantum oscillations. Increasing
$\Gamma$ broadens each level and causes neighboring levels to overlap,
thereby averaging out the discrete spectral structure. Consequently, the
oscillatory component is progressively suppressed, and the response evolves
toward a smooth background. This behavior confirms that the oscillations in
$\Delta T/\delta\mu_s$ originate from Landau quantization and are damped by
finite quasiparticle lifetime effects. 

The overlap of the opposite spin-split Landau level energies $\varepsilon_{n\sigma}= \sgn(n) \hbar \omega_c\sqrt{|n|} - J_0 S_0 \sigma/2$ enhances the quantum oscillation amplitude. The oscillation at an arbitrary energy for each spin branch can be calculated as 
\begin{align}
    \Delta(1/B)_\sigma(\varepsilon) = \frac{2e\hbar v_{\rm F}^2}{(\varepsilon + JS\sigma)^2}.
\end{align}
The transition between opposite spin branches comes from the DOS overlapping $D_+^{\rm LL}(\varepsilon)D_-^{\rm LL}(\varepsilon + \hbar \omega)$. The frequency of quantum oscillation for each spin branch is defined as
\begin{align}
    F_+(\mu) &= \frac{(\mu + JS)^2}{2e\hbar v_{\rm F}^2} \\
    F_-(\mu + \hbar \omega) &= \frac{(\mu +\hbar \omega- JS)^2}{2e\hbar v_{\rm F}^2}
\end{align}
at the vicinity of the Fermi energy. The contribution of the transition energy comes from the magnon's $\hbar \omega$. When the magnon energy is $\hbar \omega = 2JS$, both spin branches' oscillation frequency coincides, $F_+(\mu) = F_-(\mu + \hbar \omega)$. This hints that $\hbar \omega = 2JS$ is the only important energy for the enhancement of the quantum oscillation amplitude. This shows that the interfacial interaction $JS$ directly determines the magnon energy contribution to the amplitude of quantum oscillation. The only constraint lies in the magnon Green's function, where $\hbar \omega \geq \hbar g \mu_{\rm B} B$ must be satisfied. This leads to the inequality $2JS \geq \hbar g \mu_{\rm B} B$. For large $B$, the quantum oscillation might be suppressed. 

\section{Summary}
\label{sec:summary}

We have theoretically studied the spin Peltier effect in a graphene/FI heterostructure under a perpendicular magnetic field. 
Starting from a microscopic model of interfacial exchange coupling, we formulated the spin and energy currents generated by spin-flip scattering between electrons in graphene and magnons in the FI. 
By combining these currents with the heat-balance condition at the interface, we derived the temperature difference induced by spin accumulation in graphene, including the reduction due to phonon heat leakage across the junction.

We found that the spin-induced temperature difference exhibits pronounced quantum oscillations as a function of inverse magnetic field. 
These oscillations originate from Landau quantization in graphene: when opposite-spin Landau levels satisfy the energy condition for magnon-assisted spin-flip scattering, the interfacial spin-energy conversion is enhanced and the spin Peltier response shows a peak. 
The oscillatory structure is therefore a direct consequence of the discrete electronic spectrum of graphene. 

Although phonon heat conduction reduces the absolute magnitude of the observable temperature difference, it is expected to vary smoothly with magnetic field and therefore does not mask the Landau-level-induced oscillatory component. 
Our results suggest that spin Peltier measurements in graphene/FI heterostructures can provide a thermal probe of discrete electronic states in low-dimensional materials. 
They also complement previous studies of the SSE and SP in graphene-based junctions by demonstrating the corresponding reciprocal spin-caloritronic response.

\begin{acknowledgments}
This work was supported by the National Natural Science Foundation of China (NSFC) under Grant No. 12374126, by the Priority Program of Chinese Academy of Sciences under Grant No. XDB28000000, and by JSPS KAKENHI Grant Nos. JP23H01839, JP24H00322,
  and JP24K06951 from MEXT, Japan.
\end{acknowledgments}

\appendix

\section{Landau Levels of Graphene}
\label{app:LandauLevels}

In this appendix, we briefly summarize the derivation of the Landau levels in graphene. 
For the Landau gauge $\bm A = (0,Bx,0)$, the canonical momentum becomes $\pi_x = p_x$ and $\pi_y = p_y + eBx$.
We define annihilation and creation operators by
\begin{align}
    a &= \frac{\ell_B}{\sqrt{2}\hbar} (\pi_x -i\pi_y), \\
    a^\dagger &= \frac{\ell_B}{\sqrt{2}\hbar}(\pi_x + i\pi_y)
\end{align}
with the magnetic length $\ell_B$, for which the exchange relation $[a,a^\dagger] = 1$ holds.
The Hamiltonian is rewritten as
\begin{align}
    H_+ &= v_F (\sigma_x \pi_x + \sigma_y \pi_y) = \frac{\sqrt{2}\hbar v_F} {\ell_B}
    \begin{pmatrix}
        0 & a \\
        a^\dagger & 0
    \end{pmatrix}, \\
     H_- &= v_F (-\sigma_x \pi_x + \sigma_y \pi_y) =  \frac{\sqrt{2}\hbar v_F} {\ell_B}
    \begin{pmatrix}
        0 & - a^\dagger \\
    -a & 0
    \end{pmatrix} .
\end{align}
Expressing the eigenwavefunction as $\Psi({\bm r}) = (\Psi_{\rm A}({\bm r}),\Psi_{\rm B}({\bm r}))^T$, the eigenvalue equation for $H_+$ becomes 
\begin{align}
    \frac{\sqrt{2}\hbar v_F} {\ell_B} a \Psi_{\rm B}({\bm r}) &= \varepsilon \Psi_{\rm A}({\bm r}), \\
    \frac{\sqrt{2}\hbar v_F} {\ell_B} a^\dagger \Psi_{\rm A}({\bm r}) &= \varepsilon \Psi_{\rm B}({\bm r}).
\end{align}
From these equations, we obtain 
\begin{align}
\bigg(\frac{\sqrt{2}\hbar v_F} {\ell_B}\bigg)^2 a^\dagger a \Psi_{\rm B}({\bm r}) &= \epsilon^2 \Psi_{\rm B}({\bm r}), 
\end{align}
where the other component is given as $\Psi_{\rm A} = (\ell_B \epsilon/\sqrt{2}\hbar v_F) a \Psi_{\rm B}$.
The eigenenergy is independent of $q_y$ and is given with an integer $n$ as 
\begin{align}
    \varepsilon_n = \mathrm{sgn}(n)\,\hbar\omega_c \sqrt{|n|}.
\end{align}
The eigenfunciton $\Psi_{\rm B}$ is proportional to that of the harmonic oscillator, $\phi_{|n|,q_y}({\bm r})$, whose explicit form is given in Eq.~\eqref{phinky}-\eqref{phinky3}.
The other component of the eigenfunction is obtained as $\Psi_{\rm A} = a \Psi_{\rm B} = {\rm sgn}(n) \phi_{|n|-1,q_y}({\bm r})$ for $n\ne 0$ and as zero for $n=0$.
These results lead to Eq.~\eqref{Phiplus}.
A similar calculation can be performed for $H_-$, leading to Eq.~\eqref{Phiminus}.

\section{Nonequilibrium distribution functions}\label{appex:nonequiv}

For the derivation of the spin and energy currents, 
we also need to introduce the lesser Green's functions:
\begin{align}
\chi^<_{\rm loc}(\omega) &= \int dt \, e^{i\omega t} \chi^<_{\rm loc}(t), \\
\chi^<_{\rm loc}(t) &=
\frac{i}{\hbar} \sum_{\alpha} \langle s_{{\bm r}\alpha}^-(0) s_{{\bm r}\alpha}^+(t) \rangle_0 ,  \\
G^<_{\rm loc}(\omega) &= \int dt \, e^{i\omega t} G^<_{\rm loc}(t), \\
G^<_{\rm loc}(t) &=
- \frac{i}{\hbar} \langle S_{{\bm R}}^-(0) S_{{\bm R}}^+(t) \rangle_0 .
\end{align}
We define the nonequilibrium distribution functions in graphene and FI by
\begin{align}
f_{\rm G}(\omega) &= \frac{\chi_{\rm loc}^<(\omega)}{2i \, {\rm Im} \, \chi_{\rm loc}^R(\omega)}, \label{fGdef} \\
f_{\rm FI}(\omega) &= \frac{G_{\rm loc}^<(\omega)}{2i \, {\rm Im} \, G_{\rm loc}^R(\omega)} . \label{fFIdef}
\end{align}
In addition, advanced Green's functions are also needed.
They are defined by
\begin{align}
\chi^A_{\rm loc}(\omega) &= \int dt \, e^{i\omega t} \chi^A_{\rm loc}(t), \\
\chi^A_{\rm loc}(t) &=
\frac{i}{\hbar} \theta(-t) \sum_{\alpha} \langle [s_{{\bm r}\alpha}^+(t),s_{{\bm r}\alpha}^-(0)] \rangle_0 , \notag \\
G^A_{\rm loc}(\omega) &= \int dt \, e^{i\omega t} G^A_{\rm loc}(t), \\
G^A_{\rm loc}(t) &=
\frac{i}{\hbar} \theta(-t)  \langle [S_{{\bm R}}^+(t),S_{{\bm R}}^-(0)] \rangle_0 , 
\end{align}
We note that $\chi^A_{\rm loc}(\omega) = (\chi^R_{\rm loc}(\omega))^*$ and $G^A_{\rm loc}(\omega) = (G^R_{\rm loc}(\omega))^*$ 
hold.

\begin{widetext}

\section{Detailed Derivation of the Spin Current}\label{appex:derivationofSpinCurrent}

Here, we derive Eq.~(\ref{eq:spincurrent}) from the contour-ordered Keldysh expression~\cite{Bruus2004,Kato2019}.
\begin{align}
j_{\rm S}
&= -\frac{1}{\cal A} \, \text{Im}\,\bigg[ \sum_{{\bm r}\alpha} \, J_{ {\bm r}\alpha} \bigg\langle S_{\bm{r}}^+(\tau_1)s_{{\bm r}\alpha}^-(\tau_2) \exp\bigg(-\frac{i}{\hbar}\int_C d\tau H_{\text{int}}(\tau) \bigg)\bigg\rangle_0\bigg] \notag \\
&= -\frac{1}{2\hbar {\cal A}} \, \text{Re}\,\bigg[ \sum_{{\bm r}\alpha} \sum_{\bm{r}^{\prime}\alpha'} J_{{\bm r}\alpha} J_{\bm r' \alpha'} \int_C d\tau \, \bigg\langle \mathcal{T}_{\rm K} S_{\bm{r}}^+(\tau_1) s_{{\bm r}\alpha}^-(\tau_2) S_{ \bm{r}^{\prime}}^- s_{{\bm r}'\alpha'}^+(\tau)\bigg\rangle_0\bigg] ,
\end{align}
where $\mathcal{T}_{\rm K}$ is the contour-time ordering operator, $A(\tau) = e^{\tau H_0/\hbar} A e^{-\tau H_0/\hbar}$ denotes contour-time evolution under $H_0 = H_{\rm F} + H_{\rm G}$, $\langle \cdots \rangle_0$ is the thermal average with respect to $H_0$, and $\tau_1=(t,+)$, $\tau_2=(t,-)$ are contour times.
Using Eq.~\eqref{eq:Jav}, 
\begin{align}
j_{\rm S} &= -\frac{\hbar J_{\rm int}^2}{2{\cal A} }\,\text{Re}\,\bigg[ \sum_{{\bm r}} \int_C d\tau \, G_{\rm loc} (\tau_1,\tau) \chi_{\rm loc}(\tau,\tau_2)\bigg] \notag \\
&= - \frac{J_{\rm int}^2 n_{\rm int}}{2} \,\text{Re}\, \bigg[ \int_{-\infty}^{\infty} \frac{d(\hbar \omega)}{2\pi} \, \bigg(\chi^{<}_{\rm loc}(\omega)G^{R}_{\rm loc}(\omega) + \chi^{A}_{\rm loc}(\omega) G^{<}_{\rm loc}(\omega)\bigg)\bigg],
\end{align}
where the second line follows from the Langreth rules~\cite{Stefanucci2013}.
Finally, substituting Eqs.~\eqref{fGdef} and \eqref{fFIdef} yields Eq.~(\ref{eq:spincurrent}). 

\end{widetext}

\section{Spin chemical potential in the FI}
\label{app:mus}

The spin relaxation in the FI is described phenomenologically by
\begin{align}
    j_{\rm S}=g_s^{\rm FI}\mu_{\rm FI},
\end{align}
where $g_s^{\rm FI}$ represents the strength of relaxation in the FI.
Solving this equation, together with the energy balance $j_{\rm E}=-G_{\rm ph}\Delta T$, we obtain
\begin{align}
    \frac{\Delta T/T}{\mu_{\rm G}}
    =
    -\frac{
    L_{ES}g_s^{\rm FI}
    }{
    \left(L_{EE}+G_{\rm ph}T\right)
    \left(L_{SS}+g_s^{\rm FI}\right)
    -
    L_{SE}L_{ES}
    }.
\end{align}
This expression reduces to Eq.~\eqref{eq:generaltempdiff} in the strong-relaxation limit $g_s^{\rm FI}\to\infty$, whereas $\Delta T\to0$ in the floating-magnon limit $g_s^{\rm FI}\to0$.

\bibliography{ref}

\end{document}